\documentclass[12pt]{article}
\usepackage[english]{babel}
\usepackage[dvips]{graphicx}
\usepackage{wrapfig}
\begin{document}
\title{Tunneling current induced phonon generation in nanostructures}

\author{ P.I. Arseyev \thanks{e-mail: ars@lpi.ru}, N.S. Maslova $^\dag$, \\
 P.N.Lebedev Physical Institute of RAS, Leninskii pr.53, 119991 Moscow, Russia \\
 $^\dag$ Department of Physics, Moscow State University, 119992 Moscow, Russia}
\date{}
\maketitle
\begin{abstract}
 We analyze generation of phonons in tunneling structures with two electron states coupled
 by electron-phonon interaction. The conditions of strong vibration excitations are
 determined and dependence of non-equilibrium phonon occupation numbers on the
applied bias is found. For high vibration excitation levels self consistent
theory for the tunneling transport is presented.
\end{abstract}



 Tunneling current induces generation of phonons (or vibrational
quanta for a molecule) leading to effective "heating" of the
phonon subsystem in any real system with electron-phonon interaction.
In scanning tunneling microscopy experiments this effect may induce
motion, dissociation or desorption of adsorbed molecules thus allows single
molecule manipulation on a surface \cite{Ei},\cite{Smit}, \cite{Lee}.
The main purpose of the present work
is to reveal the conditions for strong phonon generation as well as for its suppression.
We also investigate the influence of strong phonon generation on tunneling conductivity
behavior.

In our previous work \cite{amp} we analyzed modifications of the tunneling current using
a model in which electron-phonon interaction leads to
transitions between two electron levels.
In the present paper we demonstrate that in this system tunneling current can induce
strong phonon generation. The mechanism of this enhanced phonon heating
 is closely connected with
the presence of at least two electron states in a molecule or a quantum dot,
 and in a single level model,
widely discussed in literature (\cite{Pers},\cite{Tikh},\cite{Al}), this
mechanism is absent.

We consider a tunneling system, which
is described by the Hamiltonian of the following type:
\begin{eqnarray}
             \label{H}
   \hat{H} &=& \hat{H}_{dot} + \hat{H}_{tun}  +  \hat{H}_{0}
\end{eqnarray}
The part $\hat{H}_{dot}$ corresponds to a QD or a molecule with
 two localized states taking into account. Electron-phonon interaction leads to
transitions between these two states.
\begin{equation}
     \hat{H}_{dot} =  \sum_{i=1,2} \varepsilon_{i}
                 a^+_{i}a_{i} +
              g (a^+_{1}a_{2}+a^+_{2}a_{1})(b+b^+)
              + \omega_0 b^+b
\end{equation}
where $\varepsilon_{i}$ corresponds to discrete levels in quantum dot (or
two electron states in molecule) and we adopt that $\varepsilon_{1}>\varepsilon_{2}$ ,
$\omega_0$ --- optical phonon frequency
(or molecule vibrational mode)and $g$ --- is electron-phonon coupling constant.
 Tunneling transitions from the intermediate system
 are included in $\hat{H}_{tun}$
\begin{equation}
                 \label{H_tun}
  \hat{H}_{tun} =
    \sum_{{\bf p},i=1,2} T_{{\bf p},i}(c^+_{{\bf p}}a_{i} + h.c.) +
      \sum_{{\bf k},i=1,2}T_{{\bf k},i}  (c^+_{{\bf k}}a_{i} + h.c.)
\end{equation}
And free electron spectrum in left and right electrodes (${\bf k}$ and ${\bf p}$)
includes the applied bias $V$:
\begin{equation}
     \hat{H}_{0} = \sum_{{\bf k}}
     (\varepsilon_k-\mu)c^+_{{\bf k}}c_{{\bf k}} + \sum_{{\bf p}}
     (\varepsilon_p-\mu-eV)c^+_{{\bf p}}c_{{\bf p}}
\end{equation}
Operators $c_{{\bf k}}, c_{{\bf p}}$ correspond to electrons in the leads and
$a_{i}$ - to electrons at the localized states of intermediate system with energy
$\varepsilon_{i}$.

By means of Keldysh diagram technique non-equilibrium phonon numbers can be found
from Dyson equations for phonon Green function. To determine phonon
occupation numbers we need to solve the Dyson equation for $D^<$ together with $D^{R(A)}$:
\begin{eqnarray}
     \label{Dys}
  D^{<}(\Omega) &=& D_0^{<}(\Omega)
  +D_0^{<}(\Omega)\Pi^{A}(\Omega)D^{A}(\Omega)+
 D_0^{R}(\Omega)\Pi^{<}(\Omega)D^{A}(\Omega)+
 D_0^{R}(\Omega)\Pi^{R}(\Omega)D^{<}(\Omega)
\nonumber \\
  D^{R}(\Omega) &=& D_0^{R}(\Omega)+D_0^{R}(\Omega)\Pi^{R}(\Omega)D^{R}(\Omega)
\end{eqnarray}
where
$D_0$ is equilibrium phonon Green function:
$$
D_0^{R}(\Omega) = \frac{2\omega_0}{(\Omega+i\delta)^2-\omega_0^2}
$$
$$
D_0^{<}(\Omega) =N_0(\Omega)(D_0^{R}(\Omega)-D_0^{A}(\Omega))
$$
$N_0$ is Bose distribution function.
Polarization operators in the lowest order in electron-phonon interaction
(first order in $g^2$) are easily determined:
\begin{eqnarray}
                 \label{P2}
  \Pi^{A}(\Omega)- \Pi^{R}(\Omega)&=& -4ig^2 \int \frac{d\omega}{2\pi}\left[
  ImG_2^{A}(\omega)ImG_1^A(\omega-\Omega)(n_2(\omega)-n_1(\omega-\Omega))\right.
  \nonumber \\
  &+& \left.
 ImG_1^{A}(\omega)ImG_2^A(\omega-\Omega)(n_1(\omega)-n_2(\omega-\Omega))\right]
   \nonumber \\
\Pi^{<}(\Omega)&=& -4ig^2 \int \frac{d\omega}{2\pi}\left[n_1(\omega)(n_2(\omega-\Omega)-1)
ImG_1^{A}(\omega)ImG_2^A(\omega-\Omega)\right. \nonumber \\
&+& \left. n_2(\omega)(n_1(\omega-\Omega)-1)
ImG_2^{A}(\omega)ImG_1^A(\omega-\Omega)
\right]
\end{eqnarray}
All electron Green functions in the above expressions
are calculated with full account for tunneling
transitions. Thus electron non equilibrium filling numbers $n_1, n_2$
are determined by the tunneling
processes (neglecting electron-phonon interaction) as:
$n_i(\omega)=
(\gamma^p_{i}n_p^0(\omega) + \gamma^k_{i}n_k^0(\omega))/
\gamma_{i}$.
Tunneling rates $\gamma$ are determined as usually by the corresponding
tunneling matrix elements
$T_{k,(p),i}$ and densities of states $\nu_{k,p}$ of the leads:
$\gamma^k_{i}=T_{k,i}^2\nu_k$, $\gamma^p_{i}=T_{p,i}^2\nu_p$. Total width of
electron levels due to the tunneling coupling is denoted by
 $\gamma_{i}=\gamma^p_{i} + \gamma^k_{i}$.

From Eqs.(\ref{Dys}) one can easily derive $D^<$:

\begin{equation}
       \label{D<}
  D^{<} =\frac{\Pi^<}{\Pi^A-\Pi^R}(D^R-D^A)
\end{equation}

Equation (\ref{D<}) together with the relation
 $ D^{<}(\Omega) =N(\Omega)(D^R(\Omega)-D^A(\Omega))$ leads to the following
 nonequlibrium phonon numbers:
\begin{equation}
       \label{Nph}
  N(\Omega) =\frac{\Pi^<}{\Pi^A-\Pi^R}
\end{equation}

Polarization operator $\Pi^<$ can be easily divided into two parts:
\begin{equation}
         \label{preDN}
  \Pi^<(\Omega) =2iIm\Pi^A N_0(\Omega) +i P^<(\Omega)
\end{equation}
Which allows explicitly separate the equilibrium distribution function and nonequilibrium
occupation numbers:
\begin{equation}
         \label{Nph1}
  N(\Omega) =N_0(\Omega) +\Delta N(\Omega)
\end{equation}
 Where
\begin{equation}
     \label{DN}
\Delta N(\Omega) = \frac{P^<(\Omega)}{2Im\Pi^A}
\end{equation}
Substituting in Eq.(\ref{P2}) for $n_1, n_2$ their explicit
expressions in terms of Fermi functions in the leads $n_p$ and $n_k$ we obtain:
\begin{eqnarray}
P^<(\Omega)&=&
\frac{-4g^2}{\gamma_1\gamma_2} \int d\omega
\left(n_p^0(\omega-\Omega)-n_k^0(\omega-\Omega)\right)
\nonumber \\
&\times &\left[Im G^{R}_{1}(\omega) Im G^{R}_{2}(\omega-\Omega)\left(
\gamma^k_{1} \gamma^p_{2}n_k^0(\omega)- \gamma^k_{2}
\gamma^p_{1}n_p^0(\omega)+ (\gamma^k_{1}
\gamma^p_{2}-\gamma^k_{2} \gamma^p_{1})N_0(\Omega)\right)\right.
\nonumber \\
&+&Im G^{R}_{1}(\omega-\Omega)Im G^{R}_{2}(\omega)\left. \left(
\gamma^k_{2}\gamma^p_{1}n_k^0(\omega)-
\gamma^k_{1}\gamma^p_{2}n_p^0(\omega)- (\gamma^k_{1}
\gamma^p_{2}-\gamma^k_{2} \gamma^p_{1})N_0(\Omega)\right) \right]
\end{eqnarray}
And
\begin{eqnarray}
           \label{Zn}
&& Im\Pi^A= \frac{-2g^2}{\gamma_1\gamma_2}\int d\omega
\left[ Im G^{A}_{1}(\omega) Im G^{A}_{2}(\omega-\Omega)\right. \times
\phantom{\gamma^k_{2}\gamma^p_{1}n_k^0(\omega)-
\gamma^k_{1}\gamma^p_{2}n_p^0(\omega)- (\gamma^k_{1}
\gamma^p_{2}-\gamma^k_{2} \gamma^p_{1})N_0(\Omega)}
\\
&& \left(
\gamma^k_{1} \gamma_{2}(n_k^0(\omega)-n_k^0(\omega-\Omega))+ \gamma_{2}
\gamma^p_{1}(n_p^0(\omega)-n_p^0(\omega-\Omega))+
(\gamma^k_{1}
\gamma^p_{2}-\gamma^k_{2} \gamma^p_{1})(n_k^0(\omega-\Omega)-n_p^0(\omega-\Omega))\right)
\nonumber \\
&&
+
Im G^{A}_{1}(\omega-\Omega) Im G^{A}_{2}(\omega) \times
\phantom{\gamma^k_{2}\gamma^p_{1}n_k^0(\omega)-
\gamma^k_{1}\gamma^p_{2}n_p^0(\omega)- (\gamma^k_{1}
\gamma^p_{2}-\gamma^k_{2} \gamma^p_{1})N_0(\Omega)dbfghdnjfhtygdfgr}
\nonumber \\
&&
\left.\left(
\gamma^k_{2} \gamma_{1}(n_k^0(\omega)-n_k^0(\omega-\Omega))+ \gamma_{1}
\gamma^p_{2}(n_p^0(\omega)-n_p^0(\omega-\Omega))-
(\gamma^k_{1}
\gamma^p_{2}-\gamma^k_{2} \gamma^p_{1})(n_k^0(\omega-\Omega)-n_p^0(\omega-\Omega))\right)
\right]
\nonumber
\end{eqnarray}

In the previous paper (\cite{amp}) it was shown that two different types of inelastic
tunneling current behavior exists, depending on the ratio
between four tunneling rates $\gamma^k_{1}, \gamma^p_{1}, \gamma^k_{2} \gamma^p_{2}$.
The sign of the combination $(\gamma^k_{1} \gamma^p_{2}-\gamma^k_{2} \gamma^p_{1})$
determines the relative population of the two electron levels due to the
tunneling current through the system, because the following relation holds:
$$
\frac{\gamma^k_{1} \gamma^p_{2}-\gamma^k_{2} \gamma^p_{1}}
  {(\gamma_{1} \gamma_{2})}(n_k^0(\omega)-n_p^0(\omega))
  =n_1(\omega)- n_2(\omega)
$$
Considering phonon generation processes cases of "normal" ($n_1(\omega)- n_2(\omega)<0$) and
inverse ($n_1(\omega)- n_2(\omega)>0$) population for the two levels with
$\varepsilon_1 >\varepsilon_2$ drastically differ from each other.
In the case of normal occupation phonon generation is rather weak.
Typical dependence of phonon occupation numbers
on bias voltage calculated by the help of Eq.(\ref{DN}) is depicted in Fig.1.
Maximum value of nonequilibrium phonon numbers does not exceed several units. Increasing
of temperature smears fine structure of this dependence and always decreases maximum values
of phonon generation.
\begin{wrapfigure}[17]{l}{8.1cm}
\includegraphics[width=8cm]{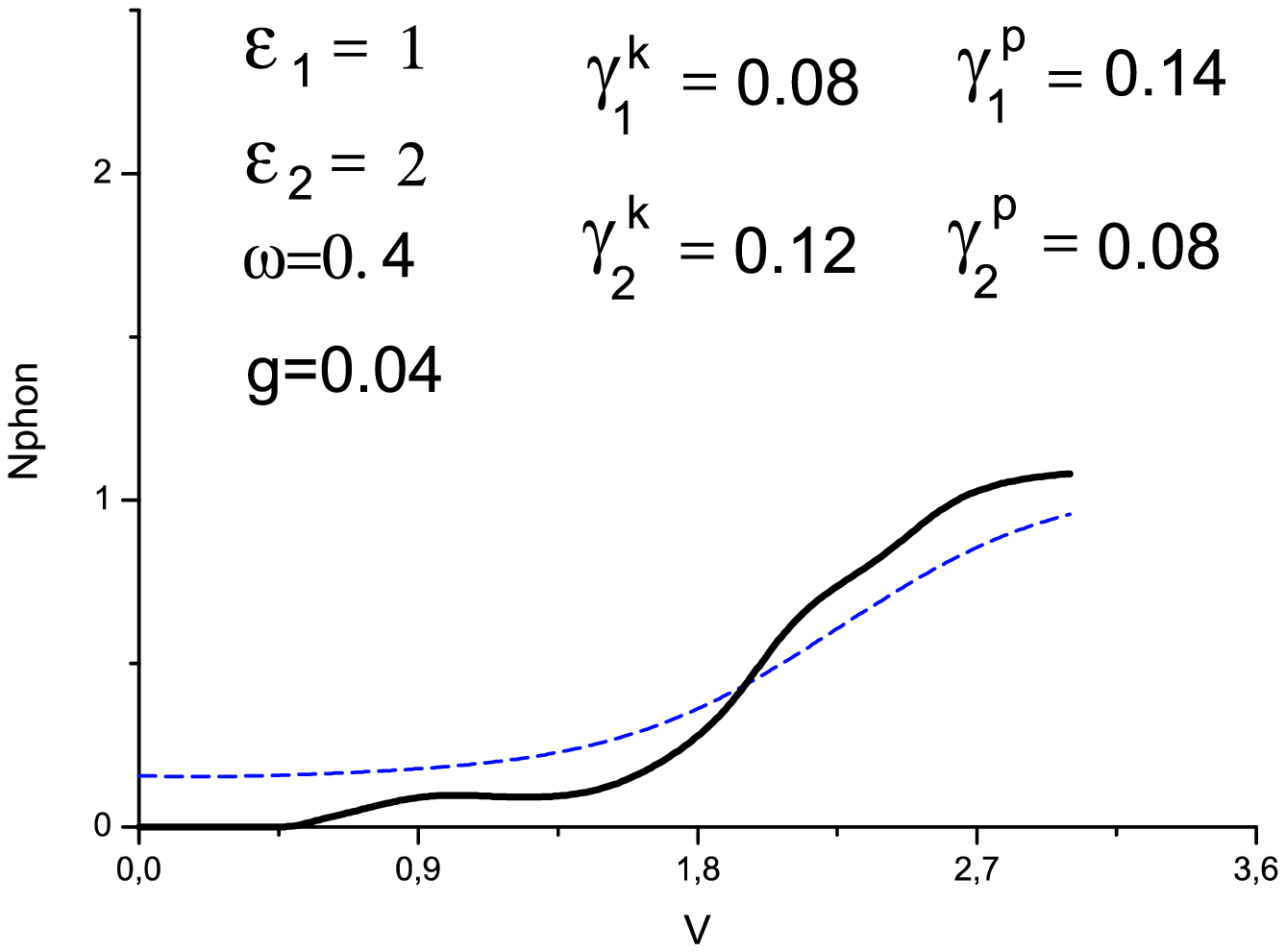}
\caption{$N_{phonon}$ versus applied bias for weak generation regime. Solid line
corresponds to zero temperature, and dashed line - to T=0.2}
\end{wrapfigure}
On the contrary
in the case of inverse occupation at some bias voltage the nonequilibrium
generation infinitly increases.
The divergence of nonequilibrium phonon filling numbers occurs because at some bias voltage
$Im\Pi^A$ given by Eq.(\ref{Zn}) passes through zero, changing its sign. In order to
describe correctly the threshold of phonon generation it is necessary to take into account
higher order terms which allows to take into account
nonlinear in phonon occupation numbers effects.
For "normal occupation" case $Im\Pi^A$ is always positive for any voltage and the
Eq.(\ref{DN}) is sufficient to calculate the non-equilibrium phonon filling numbers.

In a tunneling system with inverse electron level occupation one
should self-consistently take into account modifications of electron
Green functions due to electron-phonon interaction together with
nonequilibrium phonon numbers.
The first corrections to the electron Green functions and to the electron-phonon vertex
of the order of $g^2 D(\Omega)$ (where $D$
is dressed phonon Green function) is shown in Fig.2.
\begin{figure}[!h]
\centerline{\includegraphics[width=8cm]{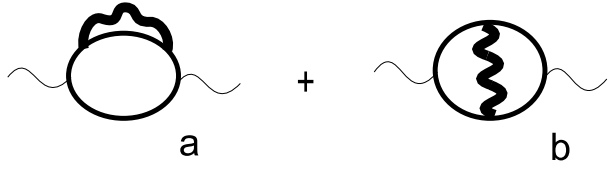}}
\caption{Diagrams for $\Pi^{(1)}$}
\end{figure}

Diagrams in Fig.2a describe the part of $Im \Pi^{(1)A}$, which is connected with the
corrections to electron Green functions. They
 can be analytically expressed by Eq.(\ref{P2}), with
$g^2$ order corrections either to $ImG^A_1$, or to $ImG^A_2$, or to occupation numbers
$n_1$ and $n_2$. Let us denote this part by $Im \Pi_g^{(1)A}$
\begin{eqnarray}
                 \label{P11}
  iIm \Pi_g^{(1)A}(\Omega)&=& -2ig^2 \int \frac{d\omega}{2\pi}\left[
  ImG_2^{(1)A}(\omega)ImG_1^A(\omega-\Omega)(n_2(\omega)-n_1(\omega-\Omega))\right.
   \\
  &+& ImG_2^{A}(\omega)ImG_1^{(1)A}(\omega-\Omega)(n_2(\omega)-n_1(\omega-\Omega))
  \nonumber \\
  &+& \left.
  ImG_2^{A}(\omega)ImG_1^A(\omega-\Omega)(n_2^{(1)}(\omega)-n_1^{(1)}(\omega-\Omega))\right]
  \nonumber \\
  &+& (1 \longleftrightarrow 2) \nonumber
\end{eqnarray}
Both the changes in electron spectral function and in occupation numbers are determined
by self-energy parts $\Sigma (\omega)$:
\begin{eqnarray}
        \label{sigma}
\Sigma^{R}_{1}(\omega)& =& i g^2\int [D^R(\omega')G^{<}_{2}(\omega-\omega')+
D^>(\omega')G^{R}_{2}(\omega-\omega')]d\omega'
\nonumber \\
\Sigma^{<}_{1}(\omega)& =& -i g^2\int D^<(\omega')G^{<}_{2}(\omega-\omega')d\omega'
\end{eqnarray}
(for $\Sigma^{R}_{2}$  $G^{<}_{2}$ is replaced by $G^{<}_{1}$.)

From the Dyson equation corrections to electron occupation
numbers are:
\begin{eqnarray}
        \label{n1}
n^{(1)}_{1}(\omega)& =& \frac{\frac{i}{2}\Sigma^{<}_{1}(\omega)+
n_1(\omega)Im\Sigma^{R}_{1}(\omega)}
{\gamma_1}
\\
n^{(1)}_{2}(\omega)& =& \frac{\frac{i}{2}\Sigma^{<}_{2}(\omega)+ n_2(\omega)Im\Sigma^{R}_{2}(\omega)}
{\gamma_2} \nonumber
\end{eqnarray}
Self-energy parts $\Sigma$ are determined by the Eqs.(\ref{sigma}), where phonon
Green functions depend on non-equilibrium phonon numbers, which should be determined
self-consistently. As we shall see below, in the weak coupling regime the width of the
phonon lines remains narrow enough even in the presence of strong phonon generation.
This allows us to use $\delta$-function
approximation for $Im D^R$ throughout all the calculations. For $n^{(1)}_{1}$ and
$n^{(1)}_{2}$ we obtain:
\begin{eqnarray}
        \label{n1fin}
n^{(1)}_{1}(\omega)& =& \frac{g^2}{\gamma_1}Im G^{A}_{22}(\omega-\omega_0)
\left[n_1(\omega)(n_2(\omega-\omega_0)-1)+N(\omega_0)(n_2(\omega-\omega_0)-n_1(\omega))
\right] - (\omega_0 \longleftrightarrow -\omega_0) \nonumber \\
n^{(1)}_{2}(\omega)& =& \frac{g^2}{\gamma_2}Im G^{A}_{11}(\omega-\omega_0)
\left[n_2(\omega)(n_1(\omega-\omega_0)-1)+N(\omega_0)(n_1(\omega-\omega_0)-n_2(\omega))
\right] - (\omega_0 \longleftrightarrow -\omega_0) \nonumber
\end{eqnarray}
It is important that some corrections appeared in $Im \Pi^{(1)A}$  are
proportional to the nonequilibrium phonon numbers $N(\omega_0)$.
These terms ensure nonlinear  limiting of phonon generation.
 In the following we retain only these terms,
because the rest part of $Im \Pi^{(1)A}$, independent of $N(\omega_0)$,
 gives only very small shift of
threshold voltage for strong phonon generation and can be omitted.

In the present problem vertex corrections to polarization operator
shown in Fig. 2b are not so small as it is
usually supposed for bulk electron phonon interaction, and also should be taken
into account.
Collecting the terms with phonon occupation numbers for diagrams in Fig.2b. we get the
second correction to $Im\Pi^A$ proportional to $N$ which we denote by $Im \Pi_v^{(1)A}$:
\begin{eqnarray}
                 \label{P12}
 && Im \Pi_v^{(1)A}(\Omega)= -2g^2 N(\omega_0)\left\{\int \frac{d\omega}{2\pi}\right.
  \times \nonumber \\
   &\times&
  Re\left[
  G_1^{A}(\omega)G_2^A(\omega-\omega_0)(G_1^<(\omega-\Omega-\omega_0)G_2^{R}(\omega-\Omega)+
  G_1^A(\omega-\Omega-\omega_0)G_2^{<}(\omega-\Omega))\right.
   \nonumber \\
  &+& (G_1^A(\omega)G_2^{<}(\omega-\omega_0)+
  G_1^<(\omega)G_2^{R}(\omega-\omega_0))G_1^{R}(\omega-\Omega-\omega_0)G_2^R(\omega-\Omega)
  \nonumber \\
  &+&
  G_1^{A}(\omega)G_2^A(\omega+\omega_0)(G_1^<(\omega-\Omega+\omega_0)G_2^{R}(\omega-\Omega)+
  G_1^A(\omega-\Omega+\omega_0)G_2^{<}(\omega-\Omega))
   \nonumber \\
  &+& \left.(G_1^A(\omega)G_2^{<}(\omega+\omega_0)+
  G_1^<(\omega)G_2^{R}(\omega+\omega_0))G_1^{R}(\omega-\Omega+\omega_0)G_2^R(\omega-\Omega)
  \right]\nonumber \\
  &+& \left.(1 \longleftrightarrow 2) \right\}
\end{eqnarray}

Integrating Eq.(\ref{Nph}) over $\Omega$ near $\omega_0$ we get the equation for
self-consistent calculation of non equilibrium phonon occupation number
$N_{ph}(\omega_0)\simeq \int d\Omega N(\Omega)
Im D^A(\Omega)$:
\begin{equation}
         \label{Nneq}
  N(\omega_0) =\frac{\Pi^<(\omega_0)}
  {2i(Im\Pi^A(\omega_0)+ Im\Pi^{(1)A}(\omega_0))}
\end{equation}
where $Im\Pi^{(1)A}=Im\Pi_g^{(1)A}+Im\Pi_v^{(1)A}$. There is no need to calculate corrections
to the function $\Pi^{<}(\omega_0)$, because it is always positive and never becomes
close to zero. So all the highest order corrections are inessential in this case.

Finally we can rewrite Eq.(\ref{Nneq}) in the following form:
\begin{equation}
         \label{Nneq1}
  N(\omega_0) =\frac{N_0(\omega_0)2Im\Pi^A(\omega_0)+P^<(\omega_0) }
  {2Im\Pi^A(\omega_0)+ g^2 N(\omega_0)A(\omega_0)}
\end{equation}
Where $P^<$ and $Im\Pi^A$ are determined by the equations (\ref{DN}) and (\ref{Zn}), and
all the essential second order corrections to $Im\Pi^A$ proportional to $N(\omega_0)$
are now rewritten as $g^2 N(\omega_0)A(\omega_0)$.
The function $A(\omega_0)$ depends on applied bias voltage through electron filling numbers
in the leads. Cumbersome expression for $A(\omega_0)$ is  presented in the Appendix for
completeness.

Equation (\ref{Nneq1}) is a simple quadratic equation with the solution:
\begin{eqnarray}
         \label{Nneq2}
 && N(\omega_0) = \frac{1}{g^2 A(\omega_0)}\left[
  -Im\Pi^A(\omega_0) \right.\nonumber \\
 &+& \left.
  \sqrt{(Im\Pi^A(\omega_0))^2+
\left(N_0(\omega_0)2Im\Pi^A(\omega_0)+P^<(\omega_0)\right)
g^2 A(\omega_0)}\right]
\end{eqnarray}

For normal electron level population (
$g^2 A(\omega_0)\ll Im\Pi^A(\omega_0)$ in all bias range)  Eq.(\ref{Nneq2}) gives only
small corrections to the first order Eq.(\ref{DN})for nonequilibrium phonon numbers.

For inverse population the appearance  of correction $g^2 A(\omega_0)$ is crucial, because
$Im\Pi^A(\omega_0)$ changes its sign and is equal to zero at some threshold
value of applied voltage.
It was found out that function $A(\omega_0)$ also changes its sign (sometimes not once)
being positive at large bias. Moreover the bias value at which $A(\omega_0)$ goes from
negative to positive values last time is very close to the threshold value for
$Im\Pi^A(\omega_0)$.
This is the reason why we should retain all the corrections of the second
order to polarization
operators in order to get the reasonable accuracy of calculations.
It was checked by direct numerical investigations, that for some parameters of the contact
only vertex corrections ensure the validity of Eq.(\ref{Nneq2}) in
all bias range.

For large voltages, greater than the threshold value, $Im\Pi^A$ becomes large negative
value. So we get from Eq.(\ref{Nneq2})the following expression for phonon occupation numbers:
\begin{equation}
         \label{Nsat}
  N(\omega_0) \simeq \frac{2|Im\Pi^A(\omega_0)|} {g^2 A(\omega_0)}
\end{equation}
For voltages beyond the threshold  and if
$|\varepsilon_1-\varepsilon_2-\omega_0|\geq \gamma_1+\gamma_2$ the value
of $Im\Pi^A(\omega_0)$ can be estimated as:
\begin{equation}
  \label{Pamax}
|Im\Pi^A(\omega_0)|\simeq
\frac{g^2}{\gamma_1\gamma_2}(\gamma^k_{1}
\gamma^p_{2}-\gamma^k_{2} \gamma^p_{1})\frac{\gamma_1+\gamma_2}
{(\varepsilon_1-\varepsilon_2-\omega_0)^2}
\end{equation}
For resonant case the denominator $(\varepsilon_1-\varepsilon_2-\omega_0)$ is replaced by
$(\gamma_1+\gamma_2)$ in this expression.
The value of $A$ in the range of large voltages can be estimated retaining only
the following resonant term of the total expression:
\begin{eqnarray}
        \label{A}
A(\omega_0)& =& -8 g^2\int \frac{d\omega}{2\pi}
(Im G^{A}_{1}(\omega))^2 (Im G^{A}_{2}(\omega-\omega_0))^2
(n_2(\omega-\omega_0)-n_1(\omega))+ ...
\end{eqnarray}
This gives for saturation regime:
\begin{equation}
A(\omega_0)\simeq
\frac{g^2}{\gamma_1\gamma_2}\frac{\gamma_1+\gamma_2}
{(\varepsilon_1-\varepsilon_2-\omega_0)^2}
\end{equation}

Therefore maximum occupation numbers in saturation regime at high voltages
are:
\begin{equation}
      \label{Nmax}
N_{max}\simeq
\frac{(\gamma^k_{1}
\gamma^p_{2}-\gamma^k_{2} \gamma^p_{1})}{g^2}
\end{equation}
Since we consider the case of weak electron-phonon interaction, $g\ll\gamma_1,\gamma_2$,
phonon occupation numbers are large, which means strong phonon generation.

Now let us return to the problem of self-consistency of the presented second order
calculations. The broadening of electron levels due to electron-phonon interaction is
determined by $Im \Sigma^A$ (\ref{sigma}). Our approximation remains valid until
$Im \Sigma^A$ is less, than the broadening of electron levels $\gamma_{1,2}$ due to the
tunneling coupling. For large phonon occupation numbers the main term of $Im \Sigma^A_1$
is:
\begin{eqnarray}
        \label{Gph}
Im \Sigma^A_1(\omega)& =& g^2N(\omega_0)
(Im G^{A}_{2}(\omega-\omega_0)+Im G^{A}_{2}(\omega+\omega_0))
+ ...
\end{eqnarray}
Thus $N(\omega_0)$ is limited by inequality:
\begin{equation}
        \label{SigmaEl}
max Im \Sigma^A_1 \simeq \frac{g^2}{\gamma_2}N(\omega_0)<\gamma_1
\end{equation}
From Eq.(\ref{Gph}) the limit of validity of our approximation is:
\begin{equation}
  \label{Nlim}
max N <\frac{\gamma_1\gamma_2}{g^2}
\end{equation}
This value of $N_{max}$ from Eq.(\ref{Nmax}) is just of the same order. So at
large values of applied bias we, strictly speaking, work at the limit of applicability
of suggested scheme but never go beyond this limit.

The phonon line width is determined by the imaginary part
of the polarization operator $\Pi^A$. Omitting second order corrections, maximum value of
$Im \Pi^A \simeq g^2/(\gamma_1+\gamma_2)$ (see (\ref{Pamax}))
is much smaller than the phonon frequency $\omega_0$.
 Second order corrections
to $Im \Pi^A$ do not change substantially this value, even for
large  nonequilibrium phonon numbers. In accordance with Eqs. (\ref{Gph},\ref{Nmax}) maximum
possible changes of electron Green functions are of the order of themselves.
Thus corrections to
polarization operators can result only in some numerical coefficient of the order of unity.
This means that even in the saturation regime with large
nonequilibrium phonon numbers the width of the
phonon line remains small.

Tunneling  current through two-level system can be expressed as \cite{amp},\cite{Win}:
\begin{equation}
        \label{Ibas}
I(V) =  2  \sum_{i,j=1,2}\gamma^k_{i}\int
  [Im G^{A}_{i}(\omega)n_k^0(\omega) - iG^{<}_{i}(\omega)]
                      d \omega
\end{equation}
Corrections to the current are thus determined by the corrections to the electron
Green functions $G^{A}_{1,2}$ and $G^{<}_{1,2}$.
First order corrections corresponding to the first diagram in Fig.3
with equilibrium phonon Green
function were calculated in the previous paper \cite{amp}. If we take into account
nonequilibrium changes of the phonon occupation numbers, we should consider four diagrams,
depicted in Fig.3.
\begin{figure}[!h]
\centerline{\includegraphics[width=12cm]{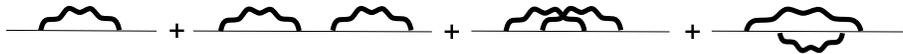}}
\caption{Corrections to electron Green function}
\end{figure}

It is worth to remark that just these types of the electron self-energy corrections
are consistent with phonon polarization operators,
depicted in Fig.2. The both sets of diagrams originate from the
the same generating functional.

This self-consistency means, that corresponding Ward identities are satisfied, and
charge conservation in the tunneling processes is automatically fulfilled.
Note that all electron Green functions
in all polarization operators and self-energy parts are calculated omitting the
electron-phonon interaction. An attempt to use self-consistent
Born approximation for the electron GF leads to violation of charge
conservation in this problem and artificial symmetrization is needed to restore
current continuity (\cite{china}, \cite{Rynd}).

Nevertheless we can neglect the contribution from the three last diagrams in Fig.3.
in the following cases:

a) If there is no inverse electron population due to the tunneling current, then phonon
numbers are small, and second order diagrams have small parameter $g^2/\gamma_1\gamma_2$.

b) If inverse population appears, but
phonon frequency is far from the resonance between electron levels:
$|\varepsilon_1-\varepsilon_2-\omega_0|\geq \gamma_1,\gamma_2$. Then
second order diagrams have small parameter $\gamma_i/|\varepsilon_1-\varepsilon_2-\omega_0|$.
In this case phonon numbers are large $N\gg 1$, so the first correction with full phonon
function strongly differs from equilibrium result.

c)If inverse population appears and
phonon frequency is almost in resonance with electron transition energy
$|\varepsilon_1-\varepsilon_2-\omega_0|\geq \gamma_1,\gamma_2$, then we can restrict
ourself to the first term only near the threshold voltage, until phonon numbers are not
too large. For high voltages in saturation regime in this case we can not use
perturbation theory, because corrections to electron Green functions become of the order
of zero order Green function.
We should sum up not only second order diagrams depicted in
Fig.3,  but all the higher order terms as well.

So we calculate the corrections to the tunneling current using only the first term in Fig.3
( one-loop diagram) but with  full nonequilibrium phonon green function, taking into
account that in resonant case c) it makes sense not far beyond the threshold of strong
generation.

In addition to the tunneling current calculated in \cite{amp} there exists a correction
to Eq. (\ref{Ibas}) proportional to nonequilibrium phonon numbers:
\begin{eqnarray}
        \label{DeltaI}
\Delta I & =& 2g^2N(\omega_0)\int \frac{d\omega}{2\pi}
\left[ \left[ \frac{\gamma^k_{1} \gamma^p_{1}}{\gamma_{1}}
Im \{(G^{(1)A}_{1}(\omega))^2\}(ReG^{A}_{2}(\omega-\omega_0)+ReG^{A}_{2}(\omega+\omega_0))
\right. \right.\nonumber \\
 &+& \left. Re \{(G^{(1)A}_{1}(\omega))^2\}
(Im G^{A}_{2}(\omega-\omega_0)+Im G^{A}_{2}(\omega+\omega_0))\right]
(n_k^0(\omega)-n_p^0(\omega)) \nonumber \\
&-& \left.
 \frac{\gamma^k_{1} \gamma^p_{2}-\gamma^k_{2} \gamma^p_{1}}{\gamma_{1}\gamma_{2}}
 (n_1(\omega)-n_2(\omega-\omega_0))Im G^{A}_{1}(\omega)Im G^{A}_{2}(\omega-\omega_0)
 + (1 \longleftrightarrow 2) \right]
\end{eqnarray}

Since $N(\omega_0)$ depends on applied bias increasing rapidly at some threshold voltage,
new peculiarities connected with $dN/dV$ can appear in the tunneling conductivity ($dI/dV$).

This peculiarity is more pronounced if $\omega_0 \ge (\varepsilon_1-\varepsilon_2)$. The
peak in $dI/dV$ at phonon generation threshold voltage (when $dN/dV \gg 1$)
is clearly seen in Fig. 4a.
\begin{figure}[!h]
\centerline{\includegraphics[width=17cm]{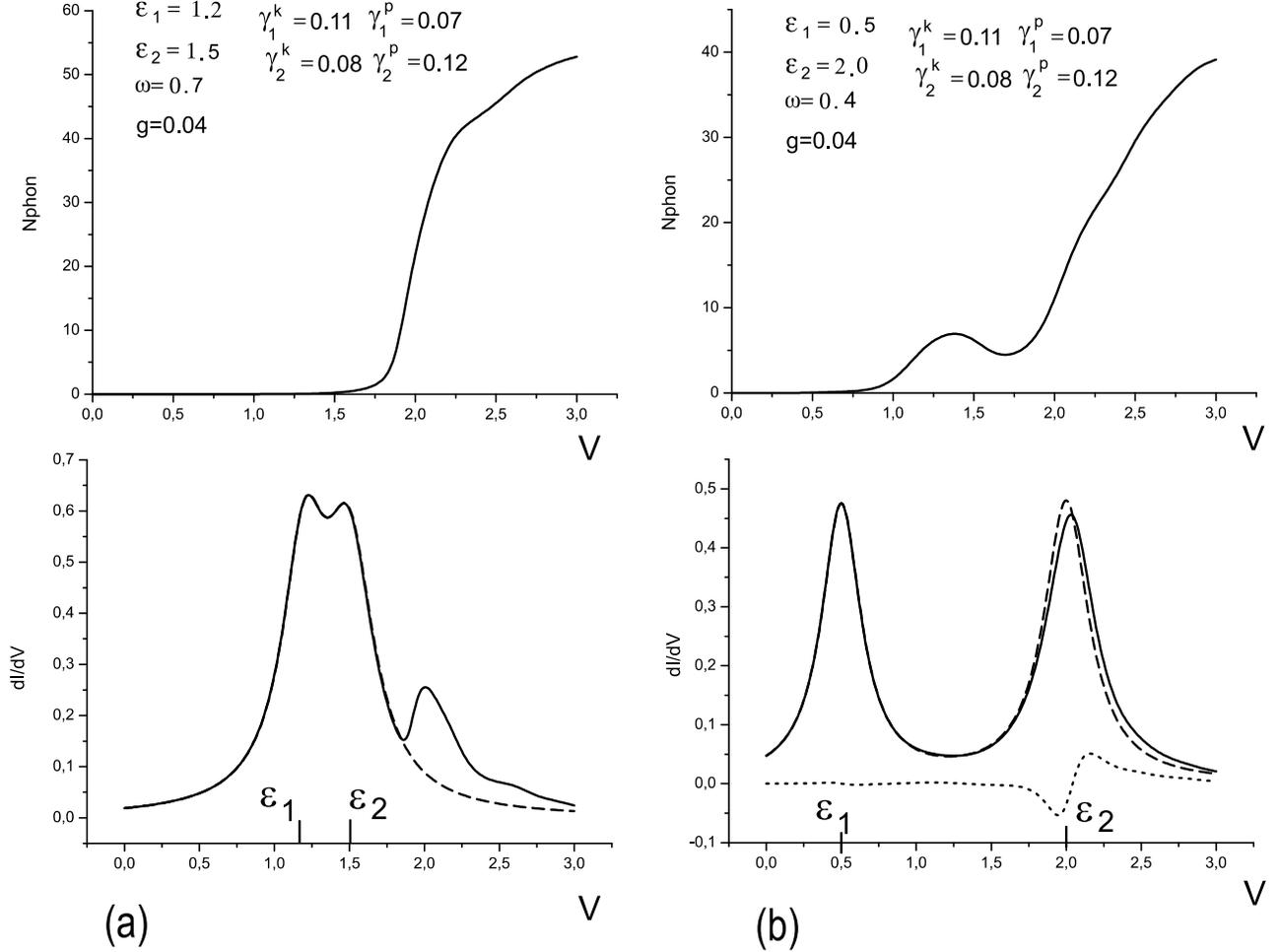}}
\caption{ $N_{phonon}$ and $dI/dV$ versus applied bias for strong
 generation regime. In the left panel (a) conductivity without electron-phonon interaction
 is shown by the dashed line. In the right panel (b) strong generation regime with
 non-monotonous behavior of $N_{phonon}$ is shown.
Correction to $dI/dV$ (dotted line) is pronounced but can not be resolved
 in the total conductivity
(solid line)
 }
\end{figure}

For $\omega_0 \le (\varepsilon_1-\varepsilon_2)$ nonequilibrium phonon numbers can depend
on applied bias non-monotonically (Fig.4b.). But in this case, in spite of great
phonon excitation, no definite peculiarities connected with electron-phonon
interaction can be observed in tunneling conductivity curves (see Fig.4b.).

Approach to saturation regime for $N$ at large bias is clearly seen in Fig.4.
The value of $N_{max}$ is in agreement with the estimation given
by Eqs.(\ref{Nmax},\ref{Nlim}).
The obtained values for $N_{max} \simeq 30-40$ for reasonable junction parameters
can hardly be observed in real small systems,
because such great overheating should lead to dissociation, desorption of molecules or
destruction of quantum dots. In order to deal with real objects at such great intensity
of vibration excitation one should take into account
phonon anharmonicity and other relaxation processes for phonons.

In conclusion we point out that intensity of phonon generation can be tuned
by changing the parameters of the tunneling junction (which influence the tunneling rates).
For inverse population of two-level electron system strong phonon generation takes place
which can lead to drastic changes of the properties of nanostructures.
Let us note that the problem of suppression of tunneling current induced
phonon generation is very important for fabrication of semiconductor cascade lasers based
on sequence of tunneling junctions \cite{laser}.
Generation of optical radiation requires the inverse
population of two electron states. But, as we have shown in the present work, in this case
strong phonon generation inevitably appears and always competes with optical
radiation generation. Using the results of the present paper one could analyze
whether it is possible to achieve the threshold of optical generation before strong
phonon generation begins
changing the parameters of the tunneling system.

This research was supported by RFBR grants  and 04-02-19957, grant for
the Leading Scientific School 4464.2006.2 and RAS Program "Strongly correlated electrons
in metals, semiconductors and superconductors". Support from the Samsung Corporation
is also gratefully acknowledged.

\newpage
 \section{APPENDIX}
Complete expression for the function $A$ in Eq.(\ref{Nneq1}) is the following:

\begin{eqnarray}
&& A(\omega_0) = -4 g^2\int \frac{d\omega}{2\pi}
2 (Im G^{A}_{1}(\omega))^2 (Im G^{A}_{2}(\omega-\omega_0))^2
(n_2(\omega-\omega_0)-n_1(\omega)) \nonumber \\
&+&Im G^{A}_{1}(\omega)Im G^{A}_{2}(\omega-\omega_0)
Im G^{A}_{1}(\omega-2\omega_0)(Im G^{A}_{2}(\omega-\omega_0)-\frac{1}{\gamma_2})
(n_1(\omega-2\omega_0)-n_1(\omega))
 \nonumber \\
&+& \frac{1}{2}Im ((G^{A}_{2}(\omega-\omega_0))^2)
[ReG^{A}_{1}(\omega)+ReG^{A}_{1}(\omega-2\omega_0)]\times \nonumber \\
&\times &[Im G^{A}_{1}(\omega)(n_1(\omega)-
n_2(\omega-\omega_0))+Im G^{A}_{1}(\omega-2\omega_0)(n_2(\omega-\omega_0)-
n_1(\omega-2\omega_0))]
 \nonumber \\
&+&Im G^{A}_{1}(\omega)Im G^{A}_{2}(\omega-\omega_0)\left[Im G^{A}_{2}(\omega-\omega_0)
Im G^{A}_{1}(\omega-2\omega_0)(2n_2(\omega-\omega_0)-n_1(\omega)-n_1(\omega-2\omega_0))
\right.+ \nonumber \\
&+& \left.Im G^{A}_{2}(\omega+\omega_0)
Im G^{A}_{1}(\omega)(n_2(\omega-\omega_0)+n_2(\omega+\omega_0)-2n_1(\omega))\right]
\nonumber \\
&+& Re G^{A}_{1}(\omega)Im G^{A}_{2}(\omega-\omega_0)\left[
ReG^{A}_{2}(\omega-\omega_0)
Im G^{A}_{1}(\omega-2\omega_0)(n_1(\omega-2\omega_0)-n_2(\omega-\omega_0))
+ \right. \nonumber \\
&+& \left.
ImG^{A}_{2}(\omega+\omega_0)
Re G^{A}_{1}(\omega)(n_2(\omega+\omega_0)-n_2(\omega-\omega_0))+
ReG^{A}_{2}(\omega+\omega_0)Im G^{A}_{1}(\omega)(n_1(\omega)-n_2(\omega-\omega_0))\right]
\nonumber \\
&+& Im G^{A}_{1}(\omega)Re G^{A}_{2}(\omega-\omega_0)\left[
ReG^{A}_{2}(\omega-\omega_0)
Im G^{A}_{1}(\omega-2\omega_0)(n_1(\omega)-n_1(\omega-2\omega_0))+\right. \nonumber \\
&+& \left.
ImG^{A}_{2}(\omega-\omega_0)
Re G^{A}_{1}(\omega-2\omega_0)(n_1(\omega)-n_2(\omega-\omega_0))+
Im G^{A}_{2}(\omega+\omega_0)Re G^{A}_{1}(\omega)(n_1(\omega)-n_2(\omega+\omega_0))\right]
\nonumber \\
&+& Im G^{A}_{1}(\omega)Im G^{A}_{2}(\omega-\omega_0)\left[ReG^{A}_{2}(\omega-\omega_0)
Re G^{A}_{1}(\omega-2\omega_0)(n_1(\omega)-n_2(\omega-\omega_0))+\right. \nonumber \\
&+& \left.
ReG^{A}_{2}(\omega+\omega_0)Re G^{A}_{1}(\omega)(n_1(\omega)-n_2(\omega-\omega_0))\right]
\nonumber \\
&+& Re G^{A}_{1}(\omega)Re G^{A}_{2}(\omega-\omega_0)\left[ImG^{A}_{2}(\omega-\omega_0)
Im G^{A}_{1}(\omega-2\omega_0)(n_2(\omega-\omega_0)-n_1(\omega-2\omega_0))+
\right. \nonumber \\
&+& \left.
ImG^{A}_{2}(\omega+\omega_0)Im G^{A}_{1}(\omega)(n_2(\omega+\omega_0)-n_1(\omega))\right]
\nonumber \\
&+& (1 \longleftrightarrow 2)
\end{eqnarray}
Typical example of polarization operators behavior for the case of normal electron population
is shown in Fig.5.
\begin{figure}[!h]
\centerline{\includegraphics[width=8cm]{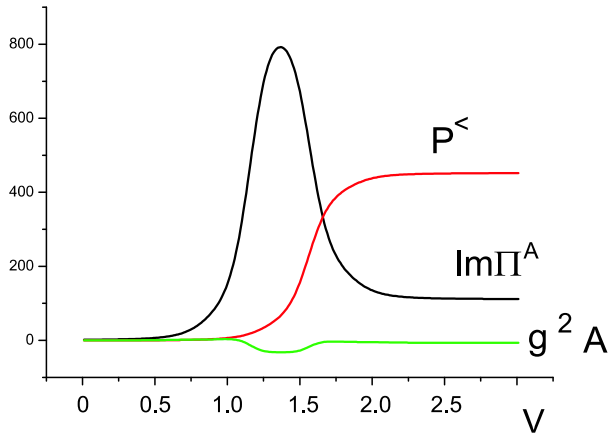}}
\caption{ Dependence of polarization operators $\Pi^<$, $Im \Pi^A$ and $A$
on applied bias for the case of normal electron population. ($\varepsilon_1=1.2,
\varepsilon_2=1.5, \omega=0.4, g=0.04,
 \gamma^k_{1}=0.08, \gamma^k_{2}=0.1, \gamma^p_{1}=0.09, \gamma^p_{2}=0.08$)
 }
\end{figure}

$Im \Pi^A$ is always positive and large enough, so corrections included
into the function $A$ are inessential. But for the inverse population we observe
quite different behavior of $Im \Pi^A$, depicted in Fig.6. If we look at the enlarged area
of the point where $Im \Pi^A=0$, we see that vertex corrections give significant contribution
to the function $A$.

\begin{figure}[!h]
\centerline{\includegraphics[width=15cm]{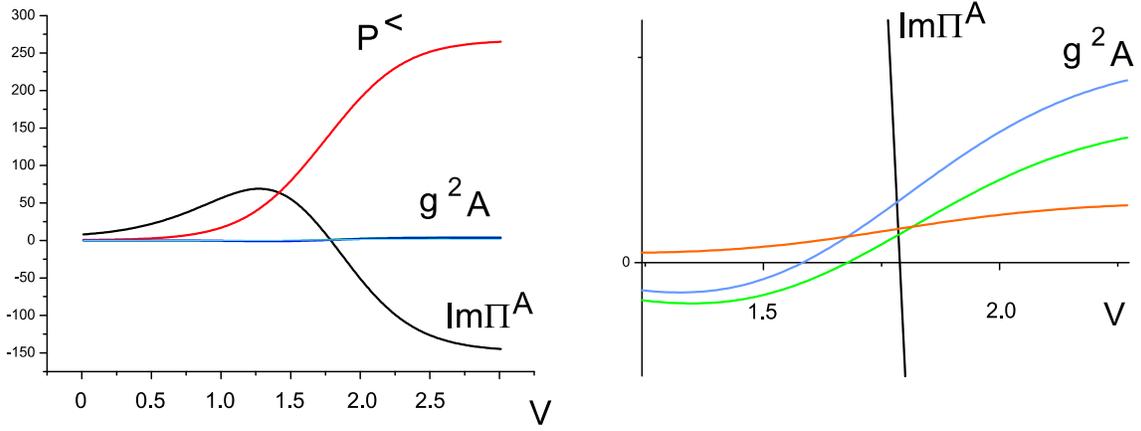}}
\caption{ The same polarization operators for the inverse population.
In the right panel enlarged
region near the bias, where $Im \Pi^A$ passes through zero, is shown. Blue line - complete
function $A$, orange line - vertex corrections, green line -corrections only to
electron Green functions
 ($\varepsilon_1=1,
\varepsilon_2=2, \omega=0.5, g=0.04,
 \gamma^k_{1}=0.14, \gamma^k_{2}=0.08, \gamma^p_{1}=0.06, \gamma^p_{2}=0.12$)
 }
\end{figure}
\end{document}